\documentclass[aps,prl,twocolumn,superscriptaddress]{revtex4-2}

\usepackage{amsmath}
\usepackage{graphicx}
\usepackage{multirow}

\usepackage{hyperref}

\usepackage{xcolor}

\begin{document}

\title{Spin-polarization of the neutron ocean in magnetar crusts}

\author{J. Servais}
\affiliation{Institute of Astronomy and Astrophysics, Universit\'e Libre de Bruxelles, CP 226, Boulevard du Triomphe, B-1050 Brussels, Belgium}
\affiliation{Brussels Laboratory of the Universe (BLU-ULB), Belgium}

\author{N. Chamel}
\email[]{nicolas.chamel@ulb.be}
\affiliation{Institute of Astronomy and Astrophysics, Universit\'e Libre de Bruxelles, CP 226, Boulevard du Triomphe, B-1050 Brussels, Belgium}
\affiliation{Brussels Laboratory of the Universe (BLU-ULB), Belgium}

\date{\today}

\begin{abstract}
Magnetars, neutron stars harboring the strongest magnetic fields known in the Universe, are associated with a broad range of observational phenomena, from giant flares to fast radio bursts, and have also been proposed as potential sites of heavy-element nucleosynthesis. These phenomena are thought to be closely linked to the structure and composition of magnetar crusts. The outer crust consists of fully ionized nuclei embedded in a degenerate electron gas, while the inner crust additionally contains free neutrons.
We investigate how strong magnetic fields modify magnetar crusts, accounting for the spin polarization of free neutrons. Arising from the coupling between the magnetic field and the neutron magnetic moment, neutron spin polarization lowers both the pressure and matter density marking the boundary between the outer and inner crusts for magnetic field strengths $B \gtrsim 10^{17}~\mathrm{G}$. As a consequence, the outer crust becomes thinner, the formation of the superheavy nuclei predicted in previous studies is suppressed, and the neutron ocean permeating the inner crust becomes spin-polarized. These effects may have important implications for the diverse manifestations of magnetars and for $r$-process nucleosynthesis in the ejecta of magnetar giant flares.
\end{abstract}

\maketitle

\section{Introduction}

Endowed with the strongest magnetic fields known in the Universe, magnetars form a subclass of isolated neutron stars \cite{Duncan1992}. Characterized by both persistent and transient activities, they are traditionally divided into two main categories: the anomalous X-ray pulsars and the soft gamma repeaters (SGRs). While they are predominantly observed in X-rays and $\gamma$-rays, magnetars have also been associated with the sources of some fast radio bursts (see, e.g., Ref.~\cite{Rea2026} for a recent review).

The surface magnetic fields of magnetars are typically of order $10^{14}-10^{15} ~\mathrm{G}$, as extracted from spectral analyses and timing measurements (see, e.g., Refs.~\cite{Kouveliotou1998,Tiengo2013,An2014}).
Recent observations, however, have suggested that some may harbor substantially stronger fields. Among others, pulse-phase modulations observed in some magnetars and interpreted as evidence for free precession imply the presence of internal toroidal magnetic fields of order $10^{16}~\mathrm{G}$ (see, e.g., Refs.~\cite{Makishima2024, Makishima2026}), and even higher fields have been inferred from radio observations of long-period transients. The slowly rotating pulsar PSR J0901$-$4046, with a period of $75.9~\mathrm{s}$, could thus have a surface field around $3\times10^{16}~\mathrm{G}$ \cite{Sobyanin2023}, and timing measurements of the radio transient ASKAP J1839-0756, with a period of $6.45~\mathrm{h}$, suggest an extreme surface dipolar field of intensity $\lesssim 10^{18}~\mathrm{G}$ \cite{Lee2025}. This latter value is comparable to the strongest fields that can be sustained in the interior of a magnetar, as found in numerical simulations~\cite{Lai1991,Uryu2019,Uryu2023}. 
For comparison, the strongest controllable magnetic field ever achieved in an indoor laboratory reached $1.2 \times 10^7~\mathrm{G}$ \cite{Nakamura2018}.

Material from magnetar crusts can be ejected during giant flares, outbursts of very high intensity primarily observed in $\gamma$-rays, as evidenced by observations of transient radio emission in the days and weeks following the 2004 giant flare from SGR~1806$-$20 \cite{Gelfand2007}. These baryon-loaded outflows have been recently reproduced in numerical simulations  (see, e.g., Ref.~\cite{Bransgrove2026}), and are expected to provide suitable conditions for the rapid neutron-capture ($r$-)process \cite{Lattimer1977,Arnould2007,Goriely2011}, suggesting that magnetars may contribute to the production of nuclei heavier than iron. This scenario has gained further support from observations of hard gamma-ray emission in the aftermath of the 2004 giant flare, which has been interpreted as arising from the radioactive decay of freshly synthesized nuclei~\cite{Patel2025}. Nucleosynthesis calculations have been recently performed, assuming the crustal material to be rapidly shock-heated, thereby dissociating nuclei into neutrons and protons while preserving the electron fraction inherited from the initial crust composition (see, e.g., Refs.~\cite{Cehula2024,Patel2025b}). However, the effects of the magnetic field on the equation of state and crustal composition have so far been neglected. The internal composition of magnetar crusts plays a significant role in the interpretation of various of their observational manifestations. 
Notably, their persistent thermal emission \cite{Vigano2013} is thought to originate in the crust, attributed to heat release from either electron captures by nuclei \cite{Cooper2010,Chamel2021} or crust fractures \cite{DeGrandis2020}. 
Similarly, the crustal composition is a necessary input for the modeling of the quasi-periodic oscillations observed in the tail of giant flares (see, e.g., Ref.~\cite{Nandi2016}). 

The outer crust of a magnetar, where the influence of strong magnetic fields is most pronounced, is structured in solid layers, each composed of a dense plasma of fully ionized ions arranged in a crystal lattice and surrounded by a gas of electrons. The nuclides constituting the crust become more neutron-rich with increasing depth, up to the transition with the inner crust, the so-called neutron-drip point, where the pressure is such that neutrons start to drip out of nuclei \cite{Blaschke2018}. 
The primary effect of magnetic fields on the crust composition is the quantization of free-electron motion perpendicular to the magnetic field lines (see, e.g., Ref.~\cite{Haensel2007}), as originally shown by Landau \cite{Landau1930} and Rabi \cite{Rabi1928} in the context of free electrons in metals. 
This effect has been extensively studied (see, e.g., Refs.~\cite{Chamel2012,Parmar2023,Basilico2025}). It notably stiffens the equation of state compared to the free-field case, leads to the appearance of heavier, less neutron-rich nuclides, and shifts the neutron-drip pressure $P_\mathrm{drip}$ and average mass density $\rho_\mathrm{drip}$ ~\cite{Chamel2015b}.
When the field is strong enough to confine electrons to the lowest Landau-Rabi level, typically $B \gtrsim 5.5\times10^{16}~\mathrm{G}$, $P_\mathrm{drip}$ increases quasi-linearly with the magnetic field. 
At $B\simeq10^{17}~\mathrm{G}$, much heavier nuclei with $Z\geq 50$ and $N\geq 126$ are expected to appear \cite{Sekizawa2023, Parmar2023,Basilico2025}.   
For $B\gtrsim10^{18}~\mathrm{G}$, some models even predict the existence of superheavy nuclei with $Z\geq 104$ in the bottom layers of the outer crust, where most of the crustal mass is concentrated. 
The astrophysical origin of superheavy nuclei remains an open question, and no evidence for their existence in Nature has yet been found \cite{Giuliani2019}.
In particular, a recent spectroscopic analysis of an X-ray burst from the active magnetar SGR~J1935+2154 found no evidence for superheavy nuclei from ejected crustal material \cite{Xie2026}. 
Finally, such strong magnetic fields can also modify the internal structure of atomic nuclei, shifting their binding energies and thereby further altering the stratification of the outer crust (see, e.g., Refs.~\cite{Basilico2015,Jiang2024,Jiang2025}). Compositions predicted with magnetic-field-dependent nuclear models still include very heavy nuclei in the deepest layers. 

Free neutrons in the inner crust are expected to be more strongly affected by the magnetic field than neutrons bound in nuclei because of their much lower number density. Despite carrying no electric charge, neutrons couple with the magnetic field through their intrinsic magnetic moment. This coupling produces a spin-dependent energy splitting, analogous to the Zeeman effect for free electrons in metals, favoring an antiparallel alignment of the neutron spins with the ambient magnetic field. Although it can influence the neutron-drip transition and thereby modify the composition and structure of the crust, this effect has so far not been considered. 

In this Letter, we show that this previously ignored neutron spin polarization modifies the thermodynamic properties of the neutron-drip transition and, consequently, the internal constitution of magnetar crusts.

\section{Outer crust  composition}

The outer crust is traditionally described within the cold-catalyzed matter hypothesis \citep{Harrison1958, Harrison1965}, under which matter is in its absolute ground state. 
At a given pressure $P$, the composition is obtained by minimizing the Gibbs free energy per nucleon $g$ \cite{Tondeur1971, Baym1971, Lai1991}.
The crust is structured in layers, each consisting of a one-component Coulomb crystal of nuclei with mass number $A$ and atomic number $Z$, immersed in a charge-neutralizing, nearly ideal strongly degenerate electron Fermi gas. Unlike in ordinary materials, electrons in magnetar crusts are relativistic. 

The quantization of electron motion becomes significant for magnetic-field intensities above the critical value
\begin{equation}
    B_\mathrm{cr} = \dfrac{m_e^2c^3}{e \hbar} \approx 4.414 \times 10^{13}~\mathrm{G}
\end{equation}
at which the electron cyclotron energy becomes comparable to the electron rest-mass energy, with $m_e$ the electron mass, $c$ the speed of light, $e$ the elementary electric charge, and $\hbar$ the Planck-Dirac constant. The index $\nu_\mathrm{max}$ of the highest Landau-Rabi level populated by electrons is dictated by the electron number density
\begin{equation}
  n_e = \dfrac{B_*}{2 \pi^2  \lambda_e^{3}} \sum_{\nu=0}^{\nu_\text{max}} d_\nu \sqrt{\gamma_e^2 - 1 - 2\nu B_*},
  \label{eq:electron_density}
\end{equation}
where $B_* \equiv B/B_\mathrm{cr}$, $\lambda_e = \hbar/m_ec$ is the reduced electron Compton wavelength, $d_\nu$ indicates the level degeneracy with $d_\nu = 1$ for $\nu = 0$ and $d_\nu = 2$ for $\nu \geq 1$, and $\gamma_e \equiv \mu_e / m_ec^2$ is the electron chemical potential $\mu_e$ in units of the electron rest-mass energy.

The matter pressure is given by that of an ideal electron relativistic Fermi gas with a small lattice correction due to Coulomb interactions between electrons and nuclei (see Supplemental Material \cite{SM} for more details). It depends on $\gamma_e$ and the nuclide constituting the crystal. Corrections due to electron exchange and charge screening effects are not expected to exceed a few percent in strongly magnetized matter \cite{Chamel2015c} and are therefore neglected in this work. 

Finally, because matter is electrically charge-neutral, the average mass density is given by $\rho=n_e M^\prime(A,Z)/Z$, where $ M'(A,Z)$ is the nuclear mass of the nuclide $(A,Z)$, including the rest mass of $Z$ electrons. 

\section{Transition with the inner crust}

The neutron-drip point delimiting the bottom of the outer crust is reached when the Gibbs free energy per nucleon exceeds the neutron chemical potential $\mu_n$ \citep{Chamel2015a, Chamel2015b}.
As discussed in Ref.~\cite{Chamel2015a}, the neutron-drip transition should be viewed as the point where a single free neutron appears in the whole crustal layer. This is a necessary condition for $\mu_n$ to be continuous in the crust. 

Although free neutrons in the inner crust are expected to be superfluid through the formation of Cooper pairs (spin-polarized pairs might be preferred in strong magnetic fields; see  Refs.~\cite{Yoshimura2025, Yoshimura2026}) as electrons in conventional superconductors, they are sufficiently dilute at the outer-inner crust boundary to neglect this effect. For the same reason, neutron-neutron interactions can be ignored. Band-structure corrections to $\mu_n$ induced by the interactions of free neutrons with the periodic nuclear lattice, similarly to free electrons in crystals, are typically of the order of a few tens of $\mathrm{keV}$ \cite{Chamel2007, Chamel2006} and are therefore negligible compared to the neutron rest-mass energy $m_n c^2 \approx 939.565$~MeV. As a result, the neutron-drip value of $\mu_n$ can be modeled independently of nuclear models. 

Due to their internal quark structure, neutrons possess a non-zero intrinsic magnetic moment, given by
$\mu^n_N = \kappa_n \mu_N$, where $\mu_N=e\hbar/2 m_p c$ is the nuclear magneton in CGS units with $m_p$ the proton mass, and $\kappa_n \approx -1.91304276$ is the neutron magnetic moment to nuclear magneton ratio \cite{Mohr2025}. 
In the presence of a uniform magnetic field $B$, the energy dispersion relation for an ideal, non-relativistic neutron Fermi gas is given by 
\begin{equation}
    \epsilon(k,\sigma) = m_n c^2 + \dfrac{\hbar^2 k^2}{2m_n} - \sigma \mu_N^n B,
\end{equation}
with $k$ the wave vector and $\sigma=\pm1$ the neutron spin projection in units of $\hbar/2$. The last term originates from the coupling between the neutron magnetic moment and the magnetic field, resulting in different energy relations for neutrons with spin states aligned parallel (spin-up, $\sigma=+1$) or antiparallel (spin-down, $\sigma=-1$) to the field direction. 
Since $\mu_N^n < 0$, the energy is minimized in the spin-down state, opposite to the well-known Pauli spin paramagnetism of electrons in metals. This results in a net spin polarization of the neutron gas. 

At the neutron-drip point, the neutron chemical potential coincides with the lowest-energy state and is thus given by
\begin{equation}
    \mu_n (B_*) = m_nc^2 - |\kappa_n|\dfrac{m_e^2c^2}{2m_p} B_* .
    \label{eq:neutronspin}
\end{equation}
The magnetization of free neutrons reduces $\mu_n$ by $\sim 0.1 - 10~\mathrm{MeV}$ relative to the free-field case, consequently shifting the transition between the outer and inner crusts. 

Following Refs.~\citep{Chamel2015a, Chamel2015b}, $P_\mathrm{drip}$ and $\rho_\mathrm{drip}$ can be determined from the electron chemical potential $\gamma_e$ at the neutron-drip point, given by the equilibrium condition
\begin{align}
  \gamma_e + \dfrac{4}{3} \left( \dfrac{4\pi}{3} \right)^{\!\!1/3} \!\! C_M \alpha \lambda_e n_e^{1/3} Z^{2/3} = \gamma_e^\mathrm{drip} , \label{eq:drip_equilibrium}
\end{align}
where $\alpha = e^2/\hbar c$ is the fine structure constant, $C_M<0$ the Madelung constant, and
\begin{align}
  \gamma_e^\mathrm{drip} \equiv \dfrac{-M'(A,Z)c^2 + A \mu_n(B_*)}{Z m_ec^2} + 1 .\label{eq:drip_threshold}
\end{align} 

\section{Results and discussion}

\begin{figure}
    \flushleft
    \includegraphics[width=0.93\hsize]{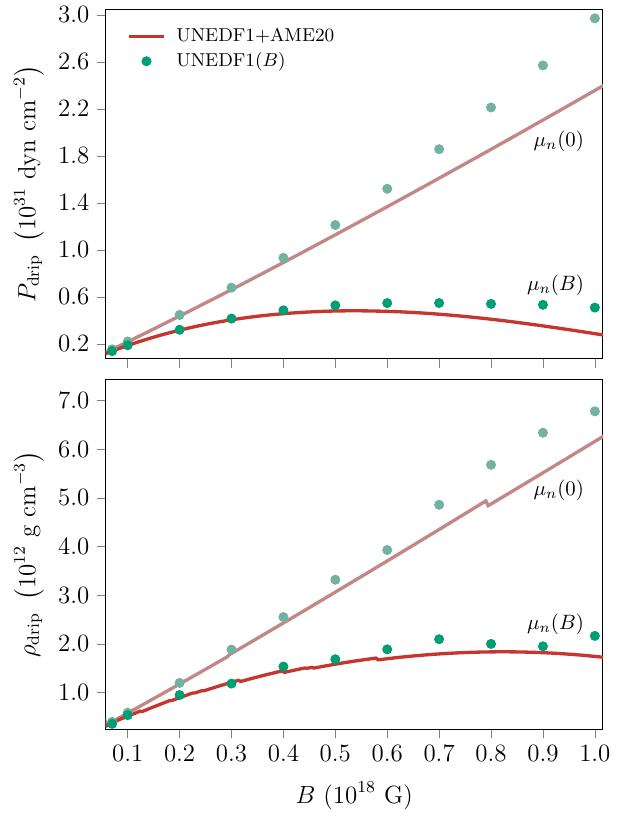}
    \caption{Matter pressure (top) and average mass density (bottom) at the outer-inner crust interface of a neutron star as a function of the magnetic field strength, with (upper curves) and without (lower curves) taking into account neutron spin polarization.}
    \label{fig:drippress}
\end{figure}

Neutron spin polarization lowers both the density and pressure at the neutron-drip point, as illustrated in Fig.~\ref{fig:drippress}. 
Numerical results were obtained with a modified version of the \texttt{magcrust} code \cite{Servais2026,Servais2026_zenodo} by computing the composition of the outer crust and extracting the properties of its deepest layer, using either experimental masses from the 2020 Atomic Mass Evaluation (AME20) \cite{Huang2021,Wang2021} supplemented with the UNEDF1 mass model \cite{Erler2012} (red lines), or the magnetic-field-dependent UNEDF1($B$) mass tables of Ref.~\cite{Jiang2025} (green dots). 
The discontinuities in $\rho_\mathrm{drip}$ reflect changes in the composition of the deepest layer of the outer crust. 
The effect of neutron spin polarization is, as expected, stronger for higher magnetic field strengths. For $B=10^{18}~\mathrm{G}$, $\rho_\mathrm{drip}$ including spin polarization is only $28\% -32\%$ of that obtained without it, while $P_\mathrm{drip}$ is reduced to only $12\%-17\%$.

The consequence of this shift for the outer crust composition of magnetars with a magnetic field strength of $5\times10^{17}~\mathrm{G}$ is shown in Fig.~\ref{fig:strat} (more examples can be seen in the Supplemental Material \cite{SM}). The sequence of equilibrium nuclides constituting the outer crust remains unchanged whether polarization is included or not, but is truncated with the polarization. This brings the neutron-drip transition closer to the stellar surface, causing the deep layers of the outer crust to disappear and extending the inner crust to lower pressures and densities. 

\begin{figure*}
    \centering
    \includegraphics[width=0.96\hsize]{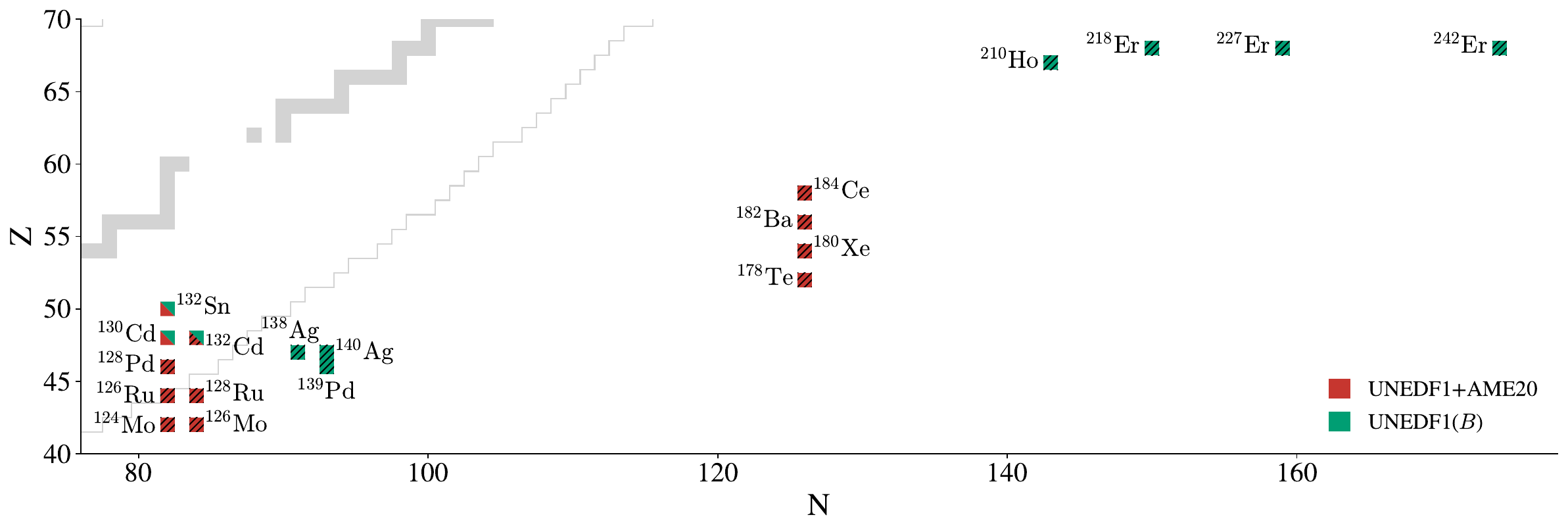} 
    \caption{Nuclide chart highlighting the nuclei constituting the deepest layers of the outer crust for $B=5\times10^{17}~\mathrm{G}$. Light-gray nuclei correspond to the valley of stability. Hatched squares indicate nuclides that disappear from the outer crust when neutron spin polarization is taken into account. Among the nuclei shown, only $^{130}$Cd and $^{132}$Sn are present irrespective of neutron spin polarization and of the nuclear model employed, whereas $^{132}$Cd remains present only for the UNEDF1($B$) mass table.}
    \label{fig:nucchart}
\end{figure*}

The nuclei constituting the outer crust are therefore less exotic than previously found since the heaviest nuclides present in the deepest layers no longer appear (see Fig.~\ref{fig:nucchart}). 
This is especially true for the superheavy nuclei predicted in Refs.~\cite{Benvenuto2023,Jiang2024,Basilico2025,Jiang2025}, as the impact of neutron spin polarization on the drip pressure becomes significant for $B\gtrsim10^{17}~\mathrm{G}$, whereas superheavy nuclei were found only for $B \gtrsim 10^{18}~\mathrm{G}$. This conclusion is unchanged considering different mass tables (see Supplemental Material \cite{SM}). These results thus suggest that magnetar outer crusts do not host superheavy nuclei, consistent with the burst spectra analysis of Ref.~\cite{Xie2026}, which reported an effective charge number of $Z\simeq37\pm14$ for the emitting plasma. 
A larger fraction of the nuclei expected in magnetar outer crusts may therefore be accessible to experimental investigations, particularly mass measurements~\cite{Yamaguchi2021}. 

\begin{figure}
    \flushleft
    \includegraphics[width=0.96\hsize]{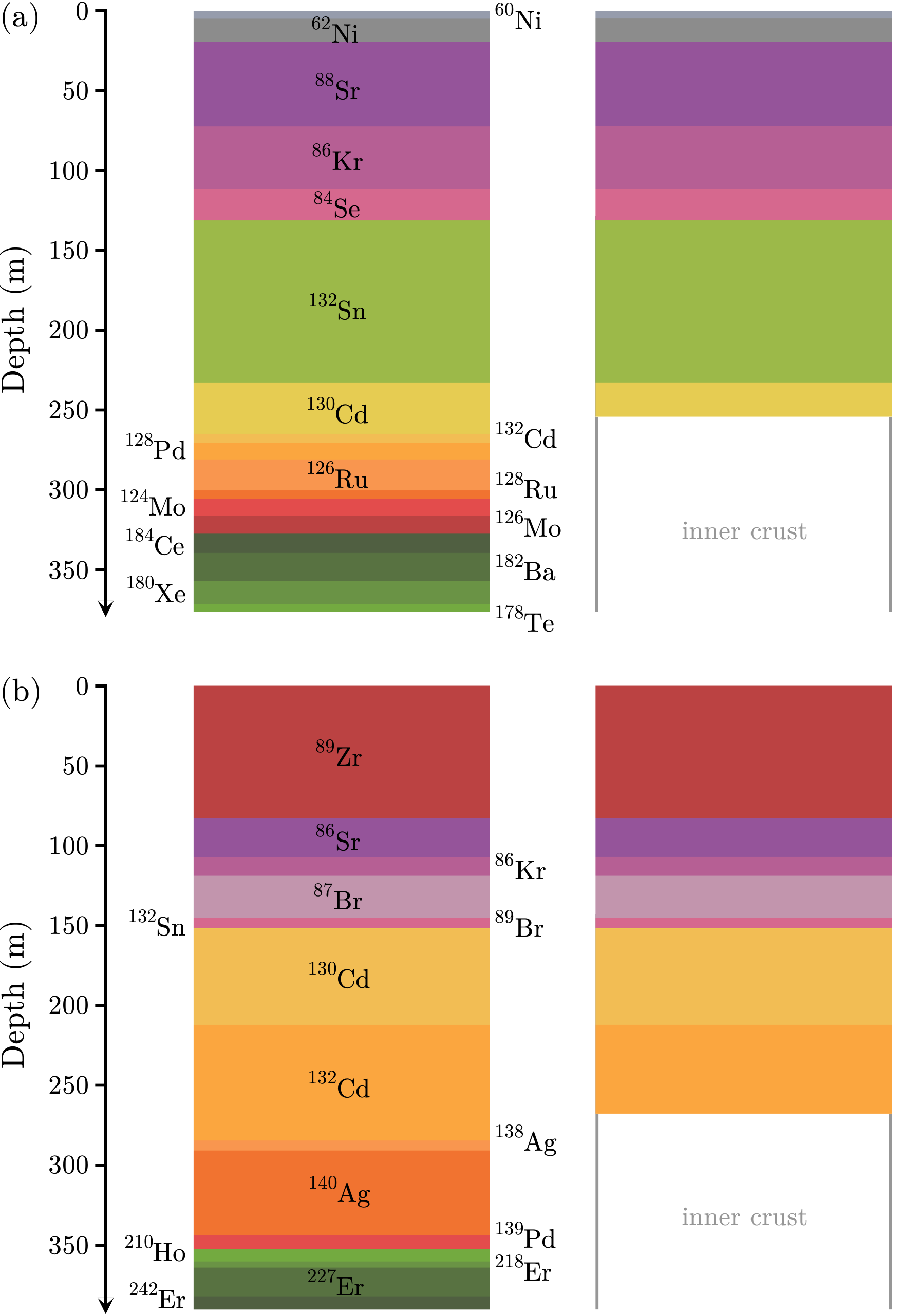} 
    \caption{Stratigraphic section of the outer crust of a magnetar with $B=5\times10^{17}~\mathrm{G}$, obtained with the mass tables UNEDF1+AME20 (a) and UNEDF1($B$) (b), with (right) and without (left) taking neutron spin polarization into account.}
    \label{fig:strat}
\end{figure}

The disappearance of the deepest layers of the outer crust may influence giant-flare $r$-process nucleosynthesis by modifying the initial composition of the ejected material.
Indeed, the electron fraction $Y_e=Z/A$ of the nuclides in the crust typically decreases with depth, with the deepest layers reaching values $Y_e \sim 0.30$ (see Supplemental Material \cite{SM}). As the deepest layers disappear because of neutron spin polarization, the electron fraction stays larger, only decreasing to $Y_e \sim 0.38-0.40$ for $B\sim10^{18}~\mathrm{G}$, thus reducing the proportion of neutrons available for capture during the $r$-process.
On the other hand, since the neutron-drip transition occurs at lower matter densities and pressures, a fraction of the inner crust could be ejected along with the outer crust. In this case, free neutrons would be present in the ejecta irrespective of the electron fraction profile, providing more favorable conditions for $r$-process nucleosynthesis. Magnetar giant flares may therefore contribute more significantly to the production of heavy elements than currently assumed.

For sufficiently strong magnetic fields, the neutron-drip transition may occur directly at the stellar surface, marking the disappearance of the outer crust. This situation is reached from a critical field strength $B_\mathrm{max}$ such that the Gibbs free energy per nucleon at the neutron-drip point coincides with the Gibbs free energy at the surface of the magnetar. An approximate analytical expression for this field is derived in the Supplemental Material \cite{SM}. It depends on the nuclide at the surface through $Z$, $A$, and its nuclear mass. Considering the surface nuclide to be $^{56}$Fe and taking its mass from AME20 yields $B_\mathrm{max}=1.58 \times 10^{18}~\mathrm{G}$. Alternatively, considering the composition obtained with the UNEDF1($B$) mass table for $B=10^{18}~\mathrm{G}$, the surface layer is made of $^{42}$Mo, and $B_\mathrm{max}= 1.84 \times 10^{18}~\mathrm{G}$. Accretion from the interstellar medium can lower $B_\mathrm{max}$ slightly as lighter elements would form the outer layer of the star, but it remains of order $10^{18}~\mathrm{G}$.  

\section{Conclusion and perspectives}

Magnetars provide unique laboratories for investigating the effects of extreme magnetic fields on dense matter. In this Letter, we have shown that neutron spin polarization, induced by the coupling of the neutron magnetic moment to the magnetic field, can substantially modify the properties of their crust for $B\gtrsim10^{17}~\mathrm{G}$. By lowering the chemical potential of free neutrons at the neutron-drip point, neutron spin polarization shifts the transition between the outer and inner crusts closer to the stellar surface. 

The consequences of this shift extend to the entire crust. 
Superheavy nuclei in the deepest layers of the outer crust predicted in previous studies now disappear, with important implications for magnetar-flare nucleosynthesis as the nuclei ejected from the crust would be less neutron-rich. 
On the other hand, the inner crust correspondingly spans a wider density range than previously thought. 
For magnetic field strengths $B \gtrsim 1.5\times10^{18}~\mathrm{G}$, the neutron ocean could even reach the stellar surface. This calls for dedicated calculations of the inner crust at densities much lower than those considered so far. 

These results could have important implications for the structure, dynamics, and observational manifestations of strongly magnetized neutron stars. For example, transport properties and elastic response of the crust may be altered, thereby influencing both the thermal evolution of the star and its oscillation spectrum, including its seismic frequencies and gravitational wave emission.
As the inner crust extends to much lower densities, the region containing superfluid neutrons could also increase. This may naturally explain why sudden changes in spin frequency, commonly attributed to the transfer of angular momentum from the inner-crust superfluid, are typically larger in magnetars than in pulsars. 

The implications of this study are furthermore not limited to magnetars. The neutron spin polarization could have particularly important consequences for strange stars and strange dwarfs, composed of a core of strange quark matter surrounded by a hadronic crust whose bottom is limited to the neutron-drip point. If such objects can sustain magnetic fields of order $10^{18}~\mathrm{G}$ (see, e.g., Ref.~\cite{Chatterjee2015}), 
the vanishing of their crust would directly expose their core with potentially observable signatures. This calls for further investigations. 

\section*{Data availability}

Full results of outer crust compositions are freely available from Ref.~\cite{Servais2026_dataset}. 

\vspace{\baselineskip}
\begin{acknowledgments}
    JS is a FRIA grantee of the Fonds de la Recherche Scientifique - FNRS (Belgium).
\end{acknowledgments}

\bibliography{paper.bib}

\end{document}

\title{{\small Supplemental Material} \\Spin-polarization of the neutron ocean in magnetar crusts}
 
 \author{J. Servais}
 \affiliation{Institute of Astronomy and Astrophysics, Universit\'e Libre de Bruxelles, CP 226, Boulevard du Triomphe, B-1050 Brussels, Belgium}
 \affiliation{Brussels Laboratory of the Universe (BLU-ULB), Belgium}
  
 \author{N. Chamel}
 \affiliation{Institute of Astronomy and Astrophysics, Universit\'e Libre de Bruxelles, CP 226, Boulevard du Triomphe, B-1050 Brussels, Belgium}
 \affiliation{Brussels Laboratory of the Universe (BLU-ULB), Belgium}

\date{\today} 

\begin{abstract}
In this Supplemental Material, further details on the outer crust model and numerical calculations are given. Additional results are presented and discussed. 
\end{abstract}

\maketitle

% Use prefix for numbering of figures, equations and tables in the SM
\renewcommand{\thefigure}{S\arabic{figure}}
\renewcommand{\theequation}{S.\arabic{equation}}
\renewcommand{\thesection}{S.\Roman{section}}
\renewcommand{\thetable}{S\arabic{table}}

{\small \tableofcontents}

\section{Composition of the outer crust of magnetars}

\subsection{Model}

The equilibrium composition of the outer crust at a given matter pressure $P$ is determined by minimizing the Gibbs free energy per nucleon. Ignoring the quantum zero-point motion of ions \citep{Baiko2009}, the pressure is explicitly given by
\begin{align}
    P =& \dfrac{B_* m_e c^2}{\left(2\pi\right)^2 \lambda_e^{3}} \, \sum_{\nu = 0}^{\nu_\mathrm{max}} d_{\nu} \left(1 + 2 \nu B_*\right) \, \psi \left( \!\sqrt{\dfrac{\gamma_e^2 - 1 - 2\nu B_*}{1 + 2\nu B_*}} \right) 
       + \left( \dfrac{4\pi}{81} \right)^{\!\!1/3} C_M \alpha\hbar c \, n_e^{4/3} Z^{2/3}, \label{eq:pressure}
\end{align}
with $\psi (x) = x \sqrt{1 + x^2} - \ln \left(x + \sqrt{1 + x^2}\right)$. The outer crust compositions presented in this work were determined iteratively, using a modified version of the code presented in Ref.~\cite{Servais2026}. Figure~\ref{fig:drippress} was obtained by extracting the properties of the neutron-drip transition as computed by this code. For the UNEDF1+AME20 mass table, outer crust compositions were computed in steps of $\Delta B_*=100$ over the interval $B_*\in[1300,23000]$, corresponding to $B\in[5.7\times10^{16},1.0\times10^{18}]~\mathrm{G}$. By contrast, UNEDF1($B$) mass tables are available only for selected magnetic-field strengths. We therefore used all available mass tables for $B\geq10^{17}~\mathrm{G}$, together with the table computed at $B=7\times10^{16}~\mathrm{G}$.

\subsection{Results}

\begin{figure}
    \centering
    \includegraphics[width=0.95\textwidth]{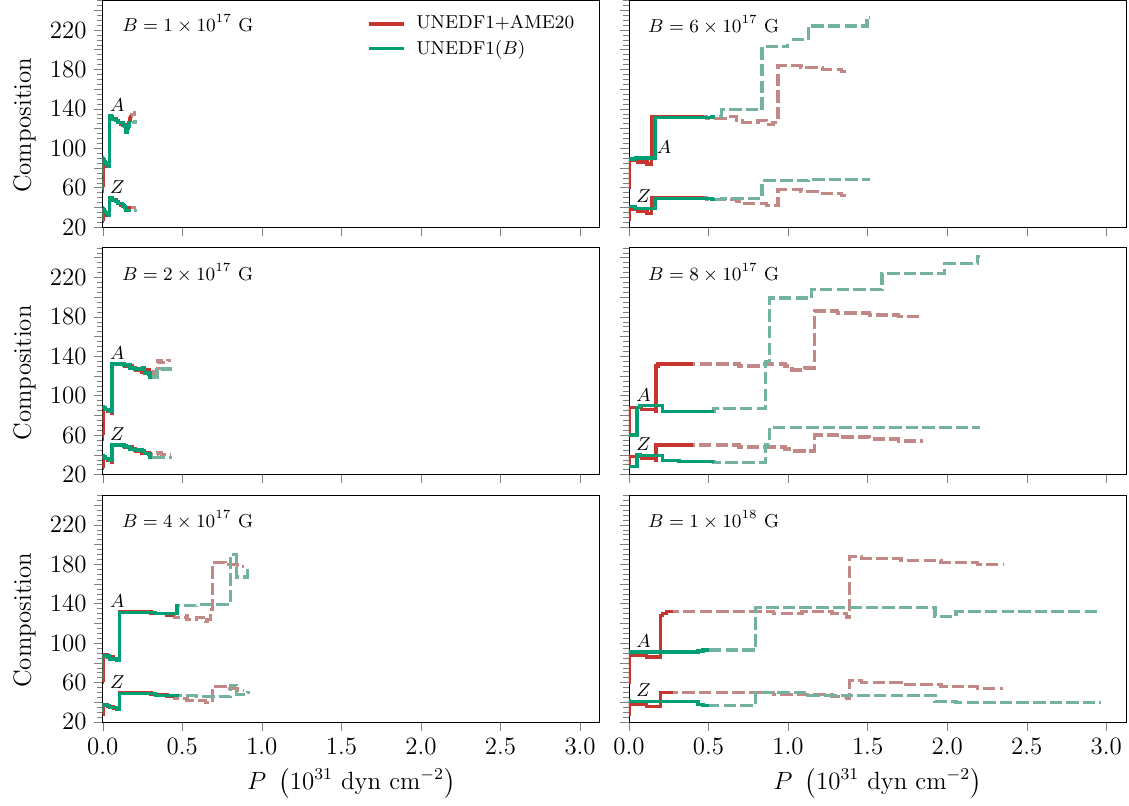}
    \caption{Outer crust composition as a function of the matter pressure for different magnetic field strengths. The regions corresponding to dashed lines disappear when taking neutron spin polarization into account.}
    \label{fig:compo_pannel}
\end{figure}

\begin{figure}
    \centering
    \includegraphics[width=0.475\textwidth]{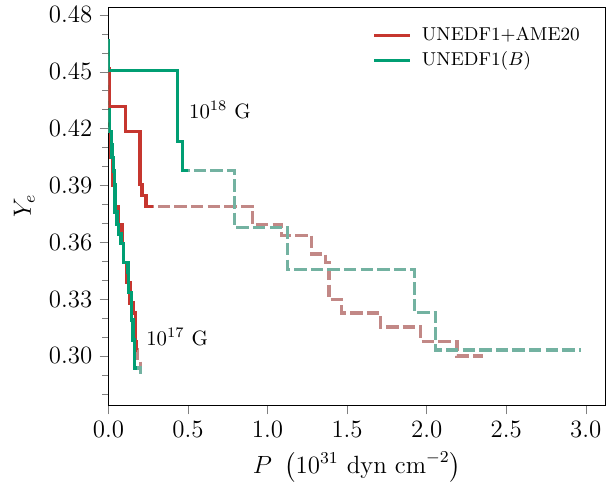} 
    \caption{Electron fraction in the outer crust of magnetars with $B=10^{17}~\mathrm{G}$ and $B=10^{18}~\mathrm{G}$. The regions corresponding to dashed lines disappear when taking neutron spin polarization into account.}
    \label{fig:elecfraction}
\end{figure}

Figure~\ref{fig:compo_pannel} illustrates the outer crust composition of magnetars with magnetic field strengths between $10^{17}~\mathrm{G}$ and $10^{18}~\mathrm{G}$.
The solid lines indicate the composition obtained when neutron spin polarization is taken into account, while the dashed extensions show the additional nuclei predicted when it is neglected. Consistent with Fig.~\ref{fig:drippress}, the outer crust extends to continuously higher pressures when neutron spin polarization is neglected, with very heavy nuclei appearing as the magnetic field increases. Considering neutron spin polarization, however, the neutron-drip pressure either stays constant (for UNEDF1($B$)) or decreases (for UNEDF1+AME20) from $B\gtrsim5\times10^{17}~\mathrm{G}$, and all nuclides in the crust satisfy $Z\leq50$ and $A<140$. The electron fraction $Y_e=Z/A$ of the nuclides in the outer crust, crucial for $r$-process simulations, is shown in Fig.~\ref{fig:elecfraction} for $B=10^{17}~\mathrm{G}$ and $10^{18}~\mathrm{G}$.

\begin{figure}
    \centering
    \includegraphics[width=0.95\textwidth]{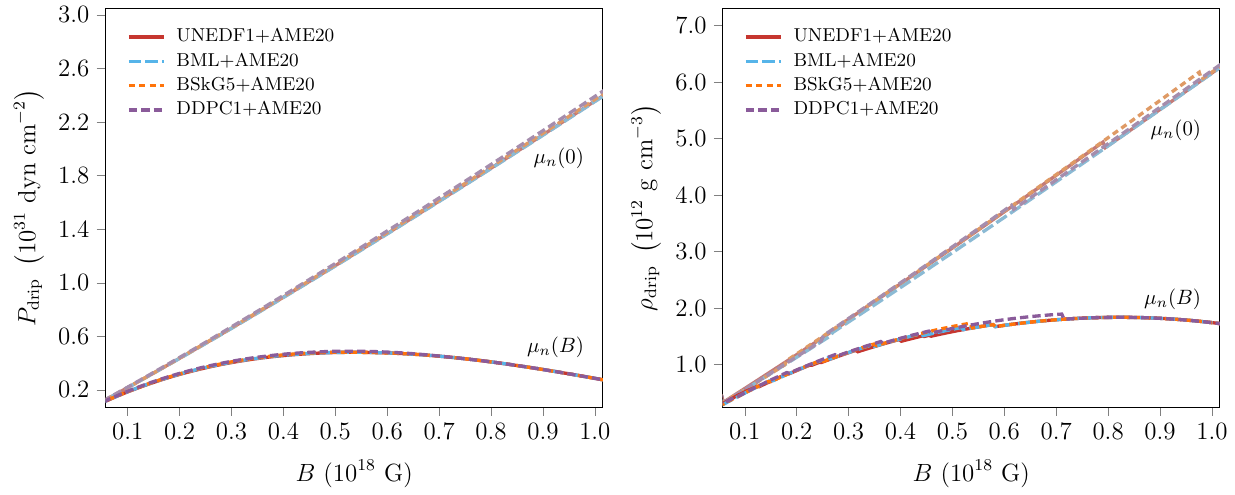}
    \caption{Matter pressure (left) and average mass density (right) at the neutron-drip transition as a function of the magnetic field strength, with (upper curves) and without (lower curves) taking into account neutron spin polarization.}
    \label{fig:drippress-multi}
\end{figure}

\begin{figure}
    \centering
    \includegraphics[width=0.95\textwidth]{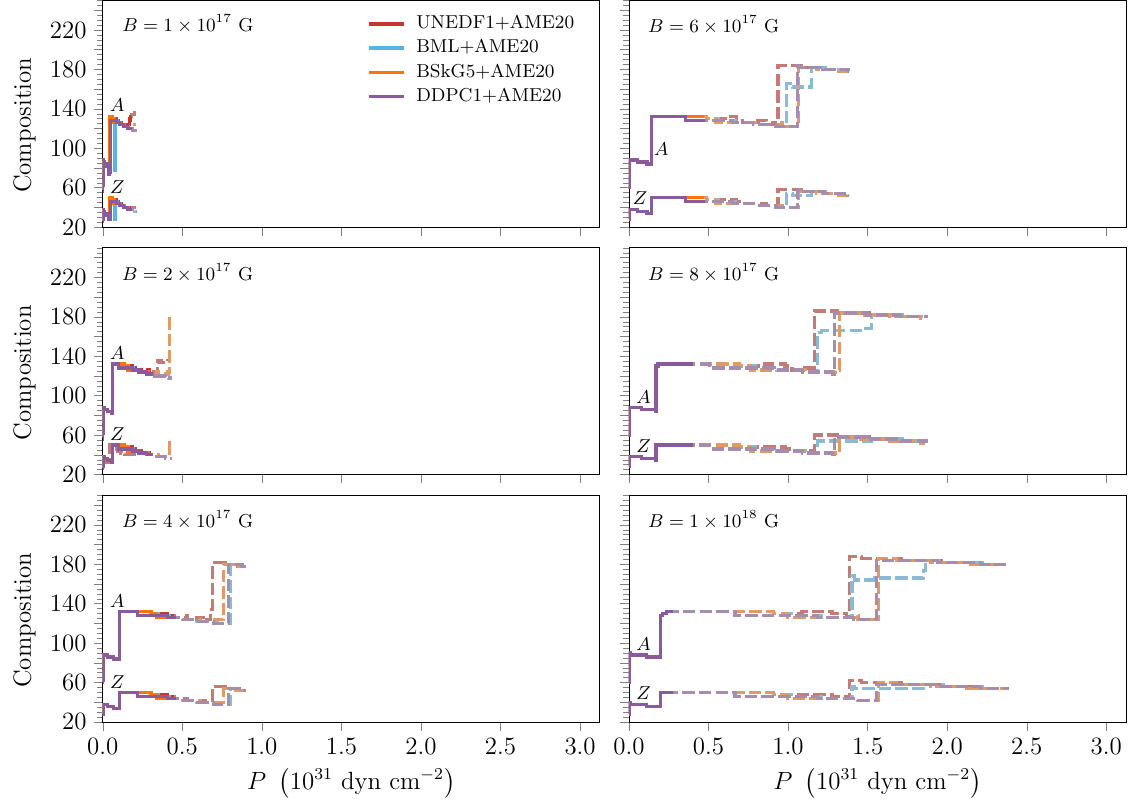}
    \caption{Outer crust composition as a function of the matter pressure for different magnetic field strengths. The regions corresponding to dashed lines disappear when taking neutron spin polarization into account.}
    \label{fig:compo_multi}
\end{figure}

To assess the reliability of our results, we tested different mass tables to supplement the AME20 data, namely BML~\cite{BML2022}, BSkG5~\cite{Grams2026} and DD-PC1~\cite{DDPC12014, DDPC12015, DDPC12016}. The neutron-drip pressure and corresponding average mass density are shown in Fig.~\ref{fig:drippress-multi}, along with the results obtained with UNEDF1+AME20 for comparison. Overall, the same behavior is observed with all mass tables. This was expected, as the neutron chemical potential at the boundary between the outer and inner parts of the crust is modeled independently of nuclear models. The results only slightly differ because the nuclides at the neutron-drip point have different masses in each model. Outer crust compositions for the same models are shown in Fig.~\ref{fig:compo_multi}. While minor model-dependent differences appear in the sequence of equilibrium nuclides, the overall compositions remain remarkably similar. The conclusions drawn above regarding the impact of neutron spin polarization on the outer and inner crusts are therefore robust against the choice of nuclear mass model.

In all cases, the obtained numerical values for the density $\rho_\mathrm{drip}$ and pressure $P_\mathrm{drip}$ were found to be very well reproduced by the approximate analytical formulas derived in Ref.~\cite{Chamel2015b} in the limits of strongly quantizing magnetic fields ($\nu_\mathrm{max}=0$) and ultra-relativistic electrons ($\gamma_e \gg 1$) using the appropriate expression for $\gamma_e^\mathrm{drip}$. They are respectively given by\footnote{A factor $1/3$ has been corrected compared to Eq.~(21) of Ref.~\cite{Chamel2015b}, originating from a missing factor $1/3$ in the lattice pressure Eq.~(9).} 
\begin{align}
    & \rho_\mathrm{drip} \approx \dfrac{M^\prime(A,Z)}{Z} \dfrac{B_*}{2 \pi^2 \lambda_e^3} \, \gamma_e^\mathrm{drip} 
    \left[ 1 - \dfrac{2 C \alpha}{3} \! \left( \dfrac{4B_*}{\pi^2} \right)^{\!\! 1/3} \! \left( \dfrac{Z}{\gamma_e^\mathrm{drip}} \right)^{\!\! 2/3} \right] \!\! , \label{eq:dripdens} \\
    & P_\mathrm{drip} \approx \dfrac{B_* m_ec^2}{4\pi^2 \lambda_e^3} \! \left( \gamma_e^\mathrm{drip} \right)^{\! 2} \! \left[ 1 - C\alpha \left( \dfrac{4B_*}{\pi^2} \right)^{\!\! 1/3} \! \left( \dfrac{Z}{\gamma_e^\mathrm{drip}} \right)^{\!\! 2/3} \right]  \!\! , \label{eq:drippress} 
\end{align}
with $C \equiv C_M \left( 4\pi / 3 \right)^{1/3} < 0$ the crystal structure constant. The analytical predictions of Eqs.~(\ref{eq:dripdens}) and (\ref{eq:drippress}) are indistinguishable from the numerical results displayed in our figures. 

\section{Critical magnetic field for the disappearance of the outer crust}
\label{app:bmax}

Accounting for the spin polarization of free neutrons at the neutron-drip point, the transition between the outer and inner crusts moves progressively closer to the stellar surface as the magnetic field strength increases. Consequently, there exists a critical magnetic field $B_\mathrm{max}$ at which the neutron-drip point reaches the stellar surface, extending the presence of free neutrons to the entire crust. Formally, this field is determined by requiring the Gibbs free energy per nucleon at the neutron-drip point, $g_\mathrm{drip}$, to coincide with the Gibbs free energy per nucleon at the surface of the star, $g_s$, as well as requiring pressure continuity such that the surface matter pressure must be zero:
\begin{align}
    & g_\mathrm{drip}(B_*) = g_s (n_e,Z_s,A_s), \label{eq:surfgibbs} \\
    & P(n_e,Z_s,B_*) = 0 \label{eq:surfpress},
\end{align}
where $(A_s,Z_s)$ is the nuclide at the surface. This condition must be solved numerically for $B_*$ to determine the critical field $B_*^\mathrm{max}$, with 
\begin{equation}
    g_\mathrm{drip}(B_*) = \mu_n(B_*),
\end{equation}
given by Eq.~\eqref{eq:neutronspin}, and (see, e.g., Ref.~\cite{Chamel2012})
\begin{align}
    g_s (n_e,Z_s,A_s) =& \dfrac{M'(A_s,Z_s)c^2}{A_s} + \dfrac{Z_s}{A_s} m_ec^2
    \left[ \gamma_e - 1 + \dfrac{4}{3} C \alpha \lambda_e n_e^{1/3} Z_s^{2/3} \right],
\end{align} 
where $\gamma_e$ and $n_e$, and potentially $M'(A_s,Z_s)$, depend on the magnetic field. 

Nevertheless, an analytical approximation for this field can be derived. Considering the field to be strongly quantizing at the surface of the star ($\nu_\mathrm{max}=0$) and that electrons at the surface satisfy $\sqrt{\gamma_e^2 -1} \ll 1$, the electron surface number density can be approximated as
\begin{equation}
    n_e^s \approx \dfrac{1}{\lambda_e^3} \left[ \left(\dfrac{B_*}{2\pi^2}\right)^{\!\! 2} |C|\alpha Z_s^{2/3} \right]^{3/5}, \label{eq:nesurf}
\end{equation}
from Eq.~\eqref{eq:surfpress} \cite{Lai1991,Chamel2012}. As a first estimate, neglecting lattice corrections, the threshold magnetic field for the appearance of free neutrons at the surface is given by 
\begin{equation}
    B_*^\mathrm{0} = \dfrac{-1}{|\kappa_n|} \dfrac{2m_p}{m_e^2c^2} \left( \dfrac{M'(Z_s,A_s)c^2}{A_s} - m_nc^2 \right). \label{eq:bmax0}
\end{equation}
A refined estimate can be obtained by adding the lattice correction to the surface Gibbs free energy, substituting $n_e$ by Eq.~\eqref{eq:nesurf}: 
\begin{align}
    g_s &= \dfrac{M'(A_s,Z_s)c^2}{A_s} 
        + \dfrac{Z_s^{5/3}}{A_s} \dfrac{4}{3} C \alpha \, m_ec^2  \left[ \left(\dfrac{B_*}{2\pi^2}\right)^{\!\!2} |C|\alpha Z_s^{2/3} \right]^{1/5} \!\!\!\!. \label{eq:musrefined}
\end{align}
Equation~\eqref{eq:surfgibbs} is then a fifth-degree polynomial equation. Now, recalling that the lattice correction is small, expressing the threshold magnetic field as 
\begin{equation}
    B_*^\mathrm{max} = B_*^0 + \delta B_* \,,
\end{equation}
where $\delta B_*/B_*^0 \ll 1$, and expanding Eq.~\eqref{eq:surfgibbs} to first order in $\delta B_*/B_*^0$, the critical magnetic field above which the outer crust vanishes is approximated as
\begin{align}
    \delta B_* &\approx - \dfrac{Z_s^{5/3}}{A_s} \dfrac{4}{3} C \alpha \, m_ec^2 \left[ \left(\dfrac{B_*^0}{2\pi^2}\right)^{\!\!2} |C|\alpha Z_s^{2/3} \right]^{1/5} 
               \left\lbrace |\kappa_n| \dfrac{m_e^2c^2}{2m_p} + \dfrac{Z_s^{5/3}}{A_s} \dfrac{8}{15} C \alpha \, m_ec^2 
               \left[ \dfrac{|C|\alpha }{(2\pi^2)^2} Z_s^{2/3} \right]^{1/5} \left( B_*^0 \right)^{-3/5} \right\rbrace^{-1} \!\!\!\!\!.
\end{align}
Taking $^{56}$Fe as the surface nuclide and using its mass from AME20, this approximation yields $B_*^\mathrm{max}\simeq 35805$, corresponding to $B \simeq 1.58 \times 10^{18}~\mathrm{G}$. This compares well with the exact numerical value $B_*^\mathrm{max} \simeq 35259$, differing by only $1.5\%$. By contrast, the zeroth-order estimate $B_*^0\simeq 34431$ differs by $2.3\%$.

\bibliography{paper.bib}